\def\be {\begin{equation}}
\def\ee {\end{equation}}
\def\ba {\begin{eqnarray}}
\def\ea {\end{eqnarray}}

\def\bea{\begin{eqnarray}}
\def\eea{\end{eqnarray}}

\documentstyle[12pt]{article}
\begin{document}

\begin{center}
{\Large \bf Microcanonical distribution for black hole \linebreak with allowance for nonadditivity}\\
\vskip .5cm
K. Ropotenko\\
\centerline{\it State Administration of Communications,}
\centerline{\it Ministry of Transport and Communications of
Ukraine,} \centerline{\it 22, Khreschatyk, 01001, Kyiv, Ukraine}
\bigskip
\verb"ro@stc.gov.ua"

\end{center}
\bigskip\bigskip

\begin{abstract}
It is shown that with allowance for nonadditivity the (canonical)
distribution for a black hole transforms to a (microcanonical)
distribution which depends on nonadditive integrals of the motion
and represents a nonlinear function of energy - the square of black
hole mass. From a comparison with the standard microcanonical
distribution it is found that the statistical weight of black hole
should be proportional to the black hole area. A definition of the
black hole entropy consistent with the statistical weight is
proposed.

\end{abstract}

\bigskip\bigskip

The statistical interpretation of the thermodynamical entropy of a
black hole
\begin{equation}
\label{ent0} S_{BH}=\frac{A}{4 l_P^{2}},
\end{equation}
where $A$ is the area of black hole horizon and $l_P$ the Planck
length, is still a central problem in black hole physics. In
statistical mechanics all properties of a system are encoded in its
distribution function \cite{lan}. For a classical system the
distribution function $\rho (q_i, p_i)$ determines the probability
to find the coordinates $q_i$ and momenta $p_i$ of the system in
intervals between $q_i$, $p_i$ and $q_i + dq_i$, $p_i + dp_i$. For a
quantum system the distribution function $w_n$ determines the
probability to find the system in a state with energy $E_n$. The
determination of this function for a system is the fundamental
problem of statistical physics. The form of the function is usually
postulated; its justification lies in the agreement between results
derived from it and the thermodynamic properties of a system.

There exist three main forms of the distribution function which play
an important part in statistical physics, corresponding in turn to a
microcanonical ensemble, whose elements have the same energy, volume
and number of particles, a canonical ensemble, with elements of same
temperature, volume and number of particles and a grand canonical
ensemble, a set of elements having equal temperature, volumes and
chemical potentials. The canonical ensemble is related to the
microcanonical ensemble in that a large system with finite
interactions can be subdivided into a relatively small independent
subsystems; the system as a whole is isolated and is described by a
microcanonical ensemble. In turn, the canonical ensemble can be
viewed as a "microcanonical" with respect to grand
canonical\footnote{In this paper we use only the microcanonical and
canonical ensembles. For this reason we shall not discuss the
interesting approach of Gour (treating  the Schwarzschild black hole
as a grand canonical ensemble) and the reader is referred to the
original paper \cite{gour}.}. Although the concept of the
distribution function is successfully used in statistical physics,
the important fact is that arguments used in deriving the
distribution function are based on the additivity properties of
ordinary matter and, more fundamentally, on the possibility of
describing a given system as made up of independent subsystems.
However the black holes are not conventional systems: they are
nonadditive. The main reason for this depart from the conventional
systems lies in the long-range behavior of gravitational forces. As
is well known, the fact that the gravitational energy is nonadditive
appears already in Newtonian gravity. In general relativity a local
definition of mass is not possible; the ADM and Komar definitions of
mass demonstrate this very clear. Moreover the black hole entropy
(\ref{ent0}) goes as the square of mass $M^{2}$ in a sharp contrast
with the additive character of entropy in ordinary thermodynamics.
It is clear that this aspect of nonadditivity can not be ignored in
deriving the distribution function for a black hole as well as in
deriving the black hole entropy. But this aspect has not yet
received attention in the literature.

In this paper I suggest the forms of distribution function and
entropy for a Schwarzschild black hole making allowance of
nonadditivity.

We begin with conventional systems. The standard determination of
the distribution function for them is given in detail by Landau and
Lifshitz \cite{lan}. Following Landau and Lifshitz consider an
ordinary isolated system consisting of quasi-isolated subsystems in
thermal equilibrium. According to Liouville's theorem the
distribution function of isolated system is an integral of the
motion. Due to the statistical independence of subsystems and, as a
consequence, multiplicativity of their distribution functions, the
logarithm of the distribution function must be not merely an
integral of the motion, but an additive integral of the motion.
There exist only seven independent additive integrals of the motion:
the energy, the three components of the momentum vector and the
three components of the angular momentum vector. It can be shown
that the statistical state of a system executing a given motion
depends only on its energy. Thus we can deduce that the logarithm of
the distribution function must be a linear function of its energy of
the form
\begin{equation}
\label{func1} \ln w^{(a)}_n = \alpha^{(a)} + \beta E^{(a)}_n,
\end{equation}
with constant coefficients $\alpha$ and $\beta$, of which $\alpha$
is the normalization constant and $\beta$ must be the same for all
subsystems in a given isolated system; the suffix "$a$" refers to
the subsystem $a$. Note that assuming another dependence $\ln w$ on
$E$ we may not obtain an additive function on the right side of
(\ref{func1}); for example, $E^{2}$ is already a nonadditive
function. Since the values of nonadditive integrals of the motion do
not affect the statistical properties of ordinary isolated system,
these properties can be described by any function which depends only
on the values of the additive integrals of the motion for the system
and which satisfies Liouville's theorem. The simplest such function
is
\begin{equation}
\label{func2} dw = const \times \delta (E -E_0)\prod \limits_a
d\Gamma_{a},
\end{equation}
where the number of states of the whole system $d\Gamma$ is a
product $d\Gamma=\prod \limits_a d\Gamma_a$ of the numbers
$d\Gamma_a$ of the subsystems (such that the sum of the energies of
the subsystems lies in the interval of energy of the whole system
$dE$). It defines the probability of finding the system in any of
the $d\Gamma$ states; $d\Gamma$ plays a role analogous to the
classical elementary volume $dq\:dp$. The factor $const$ is the
normalization constant, $\delta (E -E_0)$ the delta function. The
distribution (\ref{func2}) is called microcanonical. Note that
(\ref{func1}) is nothing but the canonical distribution if we
identify $\beta=-1/T$, $\alpha=F/T$, $F$ being the free energy and
$T$ the temperature of the system.

As is easily seen, the statistical independence and additivity play
a crucial role in deriving the distribution function for the
conventional systems. Now consider the black holes. An isolated
black hole is not stable in empty flat space. So it is helpful to
begin by placing black holes in an appropriate "box" so that a
stable equilibrium configuration of black holes and their radiation
can exist. As is well known from the thermodynamical considerations
\cite{haw}, at sufficiently high energies the entropy of such an
isolated system is dominated by a single black hole; this
corresponds to a microcanonical ensemble. As mentioned above, the
black holes are not conventional systems: they possess the
nonadditive properties. Because of the nonadditivity, the black
holes can not be thought as made up of any independent subsystems.
The presence of the logarithm in (\ref{func1}) is just determined by
the statistical independence of the subsystems. Therefore, if we
want to establish the distribution function for the black holes, we
should remove the restrictions of the statistical independence and
additivity of integrals of the motion for the subsystems of black
hole. So dropping the logarithm and suffix "a" in (\ref{func1}) we
obtain
\begin{equation}
\label{func3} w_n = f(E_n),
\end{equation}
where $f(E_n)$ represents a nonadditive integral of the motion and
is a nonlinear function of the black hole energy. The simplest such
function is the square of energy, $f(E)=\gamma E^{2}$.  Note that as
in ordinary statistics, the form of the distribution function must
be regarded only as a postulate, to be justified solely on the basis
of agreement of its predictions with the thermodynamical properties
of black holes. Our considerations are intended to make it
plausible, and nothing more. As a result, the (canonical)
distribution for a subsystem (\ref{func1}) transforms to the
(microcanonical) distribution for the whole system of the form
\begin{equation}
\label{func4} w_n = \gamma E_n^{2},
\end{equation}
where $\gamma$ is a constant coefficient. Similarly dropping the
product and suffix "a" in (\ref{func2}), we obtain
\begin{equation}
\label{func5} dw = const \times \delta (E -E_0)d\Gamma,
\end{equation}
or after integration,
\begin{equation}
\label{func6} w = const \times \Delta\Gamma,
\end{equation}
where $\Delta \Gamma$ is already the total number of states that
accessible to the whole system in a given state. The functions
(\ref{func4}) and (\ref{func6}) are obviously the same and satisfy
the same normalization condition $\sum \limits_n w_n =1$. So
comparing (\ref{func4}) and (\ref{func6}) we get
\begin{equation}
\label{func7} \Delta\Gamma \propto M^{2},
\end{equation}
where the energy of black hole is identified with its ADM mass, $M$.
Thus the statistical weight of black hole should be of the form
\begin{equation}
\label{func8} \Delta\Gamma =\frac{A}{\xi l^{2}_P},
\end{equation}
as is evident from dimensionality considerations. Here $\xi$ is a
new constant coefficient, $\xi > 1$.

Now we can define the entropy of black hole. The entropy plays a
particularly fundamental role when the microcanonical ensemble is
used. The entropy of a conventional system is given by the Boltzmann
formula
\begin{equation}
\label{ent1} S= \ln \Delta\Gamma.
\end{equation}
So assuming the usual interpretation of entropy we would have to
take a logarithm of (\ref{func8}). It appears that in this case the
generalized second law of black hole thermodynamics would be
violated \cite{ro}. The argument involves a well-known example with
the collision of black holes: two identical black holes collide,
merge, radiating gravitational wave energy, and form a third black
hole. According to (\ref{ent1}), the initial entropy of the system
is
\begin{equation}
\label{ent5} S_i=2\ln \Delta \Gamma_i= 2\ln \left(\frac{A}{\xi
l_P^{2}}\right).
\end{equation}
On the one hand, the final entropy is bounded from above by
\begin{equation}
\label{ent6} S_f=\ln \Gamma_f= \ln \left(\frac{4A}{\xi
l_P^{2}}\right).
\end{equation}
On the other hand, by virtue of the generalized second law it must
be greater then initial entropy. So we have
\begin{equation}
\label{ent7} 2\ln \left(\frac{A}{\xi l_P^{2}}\right) < S < \ln
\left(\frac{4A}{\xi l_P^{2}}\right).
\end{equation}
As is easily seen, these inequalities are satisfied only for $A <
4\xi l_P^{2}$. This means that the standard interpretation of the
entropy in terms of the logarithm of $\Delta \Gamma$ violates the
generalized second law. Thus we conclude that the statistical
interpretation of the Bekenstein-Hawking entropy is correct only if
$\log$ is deleted from the Boltzmann formula (\ref{ent1}), that is,
the black hole entropy should be proportional to the number of
states,
\begin{equation}
\label{ent2} S_{BH} \sim \Delta\Gamma.
\end{equation}
In \cite{ro} a new statistical interpretation of the black hole
entropy was suggested which does not use the properties of
additivity and statistical independence. Namely, in \cite{ro} it was
shown that the black hole entropy is of the form
\begin{equation}
\label{ent2} S_{BH} =2\pi \Delta \Gamma,
\end{equation}
that is, in contrast to the entropy of ordinary matter, without the
logarithm, and the number of microstates for a given area is $\Delta
\Gamma = A/8\pi l_P^{2}$. This agrees with our form of the
distribution function so we can identify $\xi = 8\pi$ in
(\ref{func8}). Since our distribution is determined by the
nonadditive integral of the motion (\ref{func4}), we consider the
microcanonical entropy $S_{BH}$ with fixed area rather than fixed
energy.

As is well known, a description of a black hole via a thermal
ensemble is inappropriate. Thermodynamically the heat capacity of a
Schwarzschild black hole is negative, so it cannot come to
equilibrium with an infinite heat bath. From the point of view of
statistical physics this could be explained as follows. According to
the canonical distribution, the probability $p_i$ of a system being
in a state of energy $E_i$ is proportional to the Boltzmann factor,
\begin{equation}
\label{canon1} p_i \sim \Delta \Gamma_i
\exp\left(-\frac{E_i}{T}\right),
\end{equation}
where $T$ is the temperature of the system. Assuming the usual
interpretation of entropy (\ref{ent1}), the statistical weight of
black hole should grow with $M$ as
\begin{equation}
\label{func9} \exp (4\pi GM^{2}).
\end{equation}
In that case, the total probability diverges. However, in the case
where the entropy is given by (\ref{ent2}), the statistical weight
grows as
\begin{equation}
\label{func10} 2GM^{2},
\end{equation}
and the probability can converge. So, it seems that the usual value
of the statistical weight (\ref{func9}) better explains the
breakdown of the canonical ensemble for black holes than suggested
(\ref{func10}). But this is not the case. First, it is clear that an
infinite heat bath is gravitationally unstable. On the other hand,
there is always a size of bath at which the interaction energy
between the members of ensemble is not negligible. In both cases the
the canonical distribution is not applicable. Secondly, a black hole
possesses a very special property which singles it out; namely, its
size and temperature are not independent parameters. As a result,
the temperature of black hole does not remain constant, contrary to
the definition of canonical ensemble. Note that the ordinary
self-gravitating systems, for example, stars and galaxies, also have
the negative heat capacity. And although their statistical weights
grow not so fast as (\ref{func9}), they can not be in a thermal
equilibrium with a heat bath. Thus the apparent divergence of
canonical distribution, in the case where the statistical weight
grows as (\ref{func9}), cannot be an evidence in support of the
usual interpretation of black hole entropy.

\end{document}